\documentclass[aps,prx,superscriptaddress,amsmath,amssymb,showpacs,twocolumn,notitlepage]{revtex4-1}
\usepackage{amssymb}
\usepackage{amsmath}
\usepackage{graphicx}
\usepackage{bm}
\usepackage{bbm}
\usepackage{color}
\usepackage{times}
\usepackage{xcolor}

\renewcommand{\eqref}[1]{(\ref{#1})}

\graphicspath{{./Figures/}}

\DeclareMathOperator{\Rea}{Re}
\DeclareMathOperator{\Ima}{Im}

\begin{document}

\title{Exact formulas for radiative heat transfer between planar
  bodies under arbitrary\\temperature profiles: modified asymptotics
  and sign-flip transitions}

\author{Riccardo Messina}
\affiliation{Laboratoire Charles Coulomb (L2C), UMR 5221 CNRS-Universit\'{e} de Montpellier, F- 34095 Montpellier, France}
\author{Weiliang Jin}
\affiliation{Department of Electrical Engineering, Princeton University, Princeton, NJ 08544, USA}
\author{Alejandro W. Rodriguez}
\affiliation{Department of Electrical Engineering, Princeton University, Princeton, NJ 08544, USA}

\begin{abstract}
  We derive exact analytical formulas for the radiative heat transfer
  between parallel slabs separated by vacuum and subject to arbitrary
  temperature profiles. We show that, depending on the derivatives of
  the temperature at points close to the slab--vacuum interfaces, the
  flux can exhibit one of several different asymptotic low-distance
  ($d$) behaviors, obeying either $1/d^2$, $1/d$, or logarithmic power
  laws, or approaching a constant.  Tailoring the temperature profile
  within the slabs could enable unprecedented tunability over heat
  exchange, leading for instance to sign-flip transitions (where the
  flux reverses sign) at tunable distances. Our results are relevant to the theoretical
  description of on-going experiments exploring near-field heat
  transfer at nanometric distances, where the coupling between radiative and conductive heat transfer could be at the origin of temperature gradients.
\end{abstract}


\maketitle

\section{Introduction}

Two bodies held at different temperatures and
separated by vacuum can exchange energy radiatively. At distances $d$
much smaller than the thermal wavelength $\lambda_T=\hbar c/k_B T$, such
radiative heat transfer (RHT) can be orders-of-magnitude larger than
the far-field theoretical limits predicted by Planck's law, a
consequence of evanescent tunneling~\cite{JoulainSurfSciRep05}. This
effect is further enhanced in materials supporting polaritonic
resonances, leading to a well-known divergence $\sim 1/d^2$ of the
flux with decreasing vacuum gaps~\cite{ChapuisPRB08,MuletMTE02}. Such
a divergence has been confirmed by experiments at sub-micron
scales~\cite{HuApplPhysLett08,NarayanaswamyPRB08,RousseauNaturePhoton09,ShenNanoLetters09,KralikRevSciInstrum11,OttensPRL11,vanZwolPRL12a,vanZwolPRL12b,KralikPRL12,KimNature15,StGelaisNatureNano16,SongNatureNano15},
but has been observed and predicted to fail at sub-nanometric
distances~\cite{KittelPRL05,KloppstecharXiv}. In particular,
deviations from the $1/d^{2}$ power law have been predicted to arise in
interleaved geometries~\cite{Rodriguez13}, as well as due to non-local
damping~\cite{HenkelApplPhysB06,JoulainJQSRT}, acoustic phonon
tunneling~\cite{ChiloyanNatComm15}, and from the interplay of sur-
face roughness and curvature \cite{Kruger}. One unexplored mechanism that
could potentially modify RHT are temperature variations: at nanometer
gaps (now within experimental
reach~\cite{SongNatureNano15,KloppstecharXiv}), the interplay between
RHT and conduction can produce temperature gradients within
objects~\cite{MessinaPRB,JinarXiv}, requiring full account of such effects within
the quantum-electrodynamics framework~\cite{FVC,Eda1,Eda2}.

In this work, we derive exact analytical formulas for the RHT
between two parallel slabs subject to arbitrary temperature profiles
and demonstrate the existence of several asymptotic low-distance $d$
behaviors: depending on the values and derivatives of the temperature profile at points
near the slab--vacuum interfaces, the flux can diverge as $1/d^2$,
$1/d$, or logarithmically, or approach a constant, as $d \to 0$. We
show that the temperature profile of the slabs can be tailored so as
to modify and even reverse the direction of the flux over tunable
distances. As described in~\cite{MessinaPRB}, such temperature gradients can naturally arise due to the interplay of conduction and radiation at nanometric scales, leading to constant (rather than diverging) flux rates as $d\to0$, even in the absence of phonon or non-local tunneling effects~\cite{JoulainJQSRT,ChiloyanNatComm15}. The impact of temperature profile on the properties of RHT remains so far almost unexplored. This tunability could be indeed relevant for the design of thermal devices, such as for example memories~\cite{devices} and thermal rectifiers~\cite{rect}, where the ability to tune the flux dependence on temperature and separation is very important.

\begin{figure}[t!]
\begin{center}
\includegraphics[height=4.1cm]{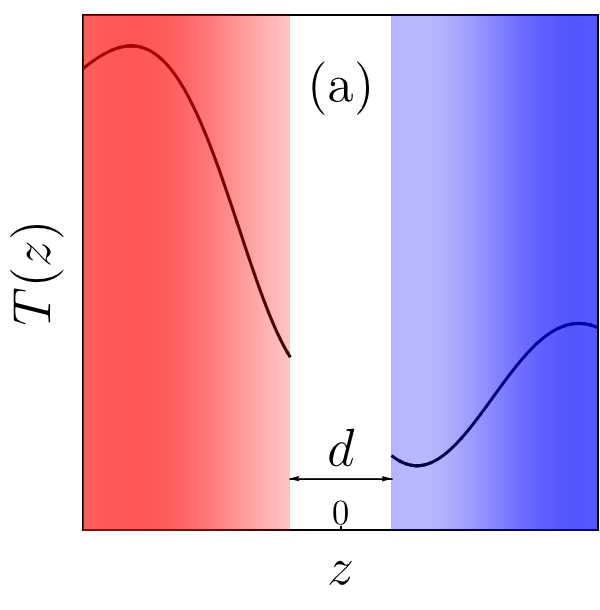}\,\,\,\includegraphics[height=4.1cm]{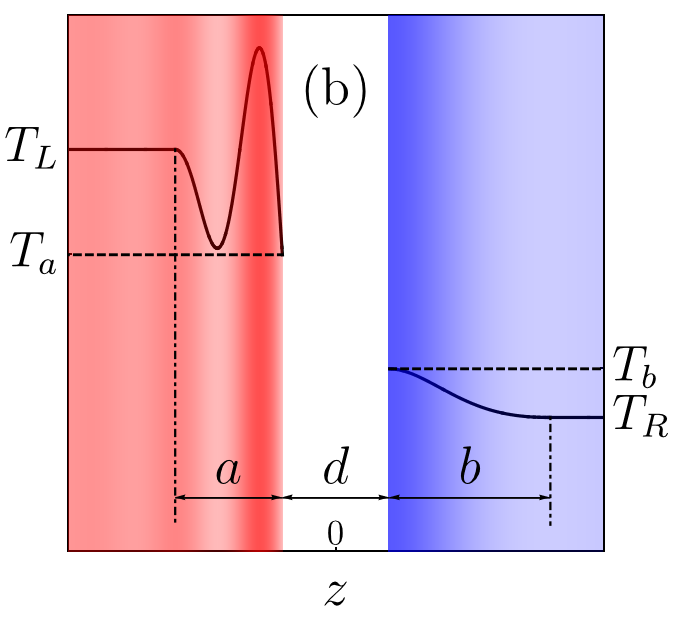}
\end{center}
\caption{Schematic of two parallel slabs separated by a distance $d$
  along the $z$ direction. The two slabs are subject to a temperature
  profile $T(z)$. Panel (a) depicts the general case of an arbitrary
  $T(z)$, whereas (b) illustrates a configuration in which the
  temperatures in the regions $z\leq -d/2-a$ and $z \geq d/2+b$ are
  held at $T_L$ and $T_R$, with $T_a$ and $T_b$ denoting the
  temperatures at the left and right slab--vacuum interfaces, respectively.}
\label{fig1}
\end{figure}

\section{General formulas}

Consider two semi-infinite co-planar slabs a
distance $d$ apart and subject to a position-dependent temperature
profile $T(z)$, represented in Fig.~\ref{fig1}(a). The RHT between the slabs is derived within the framework of the scattering-matrix formalism
developed in~\cite{MessinaPRA11,MessinaPRA14}, used previously to
describe the Casimir force and RHT in presence of two and three bodies. The first step in our derivation is to express the
correlation functions of the electric fields emitted by a single body
at temperature $T$ in terms of the reflection and transmission
operators of this body. In contrast to~\cite{MessinaPRA11}, our
scenario requires that we apply such a scheme to a film of
infinitesimally small thickness $dz$ at a position $z$ of one of the
two slabs. The total field emitted by a slab can then be calculated as
the sum of these individual fields, including contributions of
multiply reflected and transmitted fields from the other portions of
the slab, following Refs.~\cite{MessinaPRA11,MessinaPRA14}. Once the
field emitted by each slab is statistically characterized, the total
field in the vacuum gap can be deduced, allowing us to obtain the
Poynting vector or flux per unit area in the gap.

The first step in our derivation is
the characterization of the fields emitted by each body, and their
correlation functions. Assuming local thermal equilibrium, the
statistical properties of the fields radiated by each body depend only
on the local temperature within the object. Given a source of thermal
fluctuations, the quantity of interest is the symmetrized average
$\langle
E_p^\phi(\mathbf{k},\omega)E_{p'}^{\phi'\dag}(\mathbf{k}',\omega')\rangle$,
where $p$ denotes the polarization, $\phi$ the propagation direction
along the $z$ axis, $\mathbf{k}$ the component of the wavevector
orthogonal to the $z$ axis, and $\omega$ the frequency, restricted
here to positive values.

\begin{figure}[h!]
\begin{center}
\includegraphics[height=4.5cm]{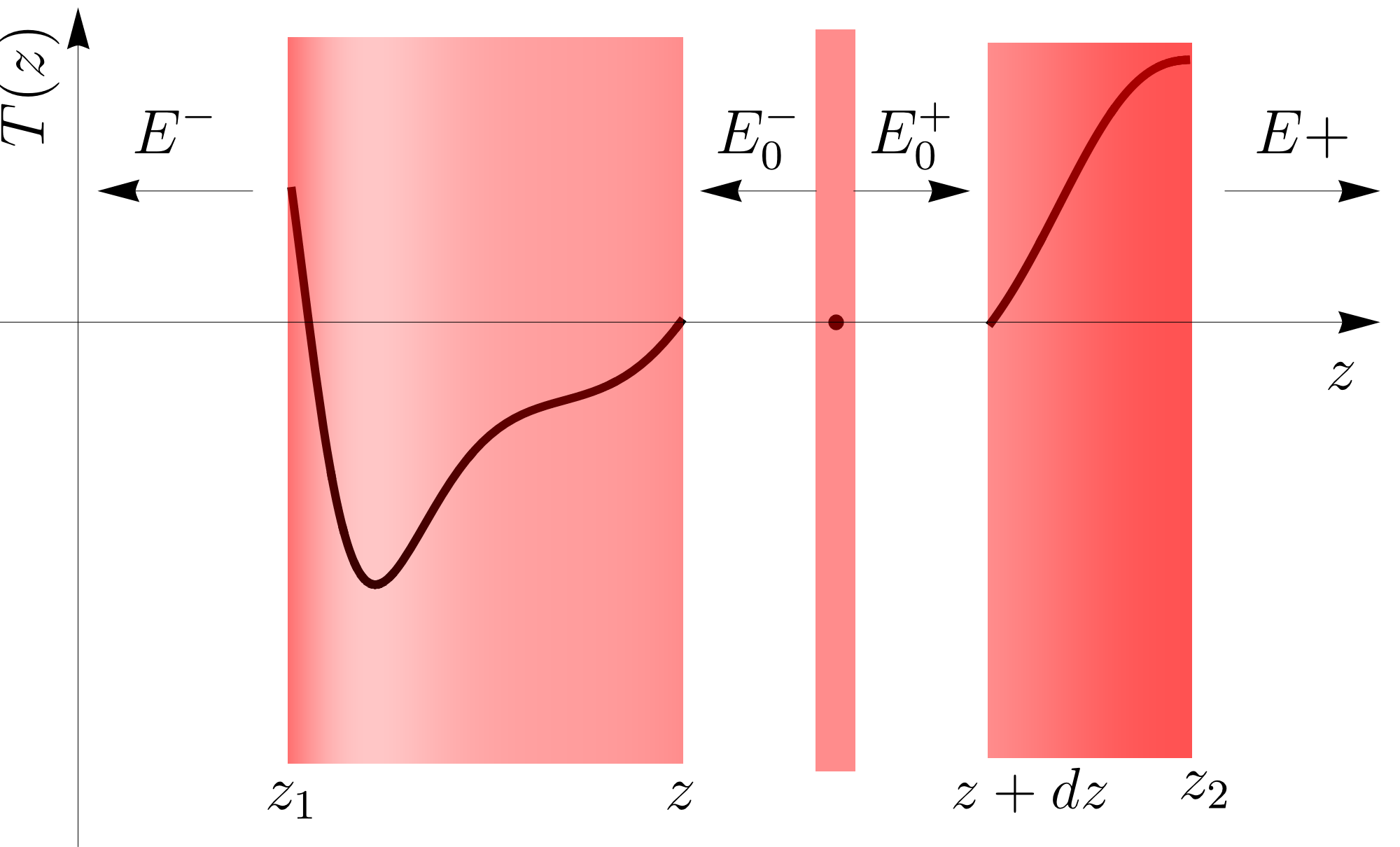}\\
\vspace{.2cm}
\includegraphics[height=4.5cm]{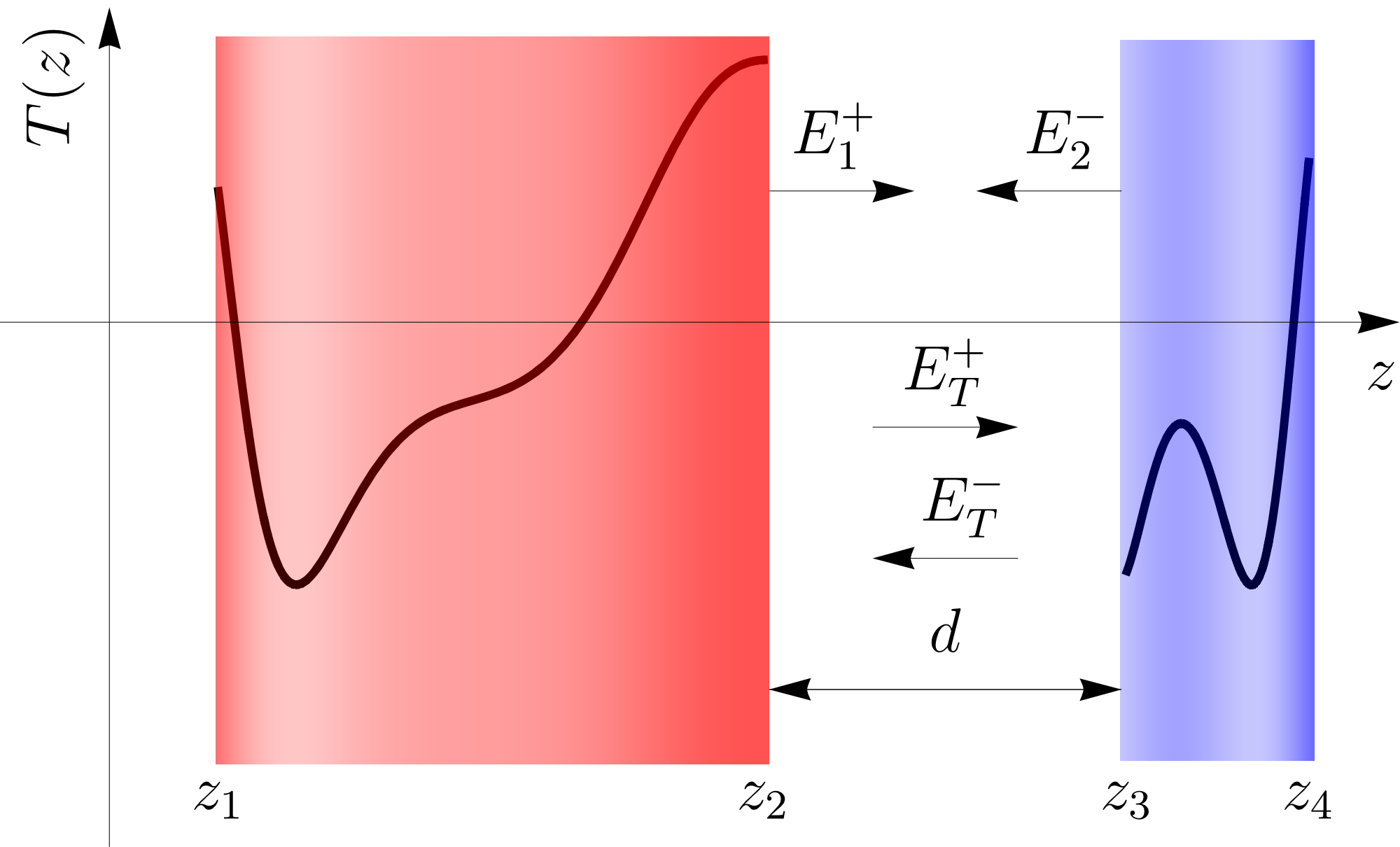}
\end{center}
\caption{In the upper part, schematic of one slab having a
  position-dependent temperature $T(z)$. The slab occupies the region
  $[z_1,z_2]$ and the element from $z$ to $z+dz$ produces the field
  $E_0$. The total field emitted by the slab on the left (right) side
  is $E^+$ ($E^-$). In the lower part, two slabs having a temperature
  profile $T(z)$ and placed at distance $d$. The slab on the left
  (right) side produces a field $E_1$ ($E_2$), while the total field
  inside the cavity has amplitudes $E_T^\pm$.}
\label{fig2}
\end{figure}

Equations (45) and (46) of \cite{MessinaPRA11} characterize the
correlation function in terms of matrix elements of the reflection and
transmission operators at each object interface. In the case of a
slab, these matrix elements coincide with the well-known Fresnel
coefficients, modified to take into account the possibility of finite
slab thickness~\cite{MessinaPRA14}. In order to incorporate the
possibility of varying temperature within a slab, we decompose the
slab in terms of infinitesimally thin films (see Fig.~\ref{fig2}) and
apply these correlation formulas to an arbitrary film located at $z$
and having thickness $dz$. Specifically, given some arbitrary position
$z$, we replace the modified Fresnel coefficients with their
first-order series expansion in terms of the thickness $dz$ of the
corresponding film, given by:
\begin{equation}\begin{split}
\rho&\simeq-i\frac{2k_{zm}r}{1-r^2}dz,\\
  \tau&\simeq1-i\Bigl(k_z-k_{zm}\frac{1+r^2}{1-r^2}\Bigr)dz,
\end{split}\end{equation}
where $k_z$ ($k_{zm}$) is the $z$ component of the wavevector in
vacuum (or the medium), and $r$ is the ordinary Fresnel
coefficient. It follows that the correlation function of the field
$E_0$ emitted by the film is given by:
\begin{equation}\label{corr}\begin{split}
 \langle
 E_{0,p}^\phi(\mathbf{k},\omega)E_{0,p'}^{\phi'\dag}(\mathbf{k}',\omega')&=\frac{\omega\,dz}{2\varepsilon_0c^2}N[\omega,T(z)](2\pi)^3\delta(\omega-\omega')\\
 &\,\times\delta(\mathbf{k}-\mathbf{k}')\delta_{pp'}\mathcal{F}_{0,p}^{\phi\phi'}(\mathbf{k},\omega).
\end{split}\end{equation}
More precisely, the field correlations involving waves traveling in
the same ($\phi'-=\phi$) or opposite ($\phi'=-\phi$) directions are
given by:
\begin{equation}\label{fpp0}\begin{split}
 \mathcal{F}_{0,p}^{\phi\phi}(\mathbf{k},\omega)&=\Theta(\omega-ck)\frac{2}{k_z}\Ima\Bigl(k_{zm}\frac{1+r^2}{1-r^2}\Bigr)\\
 &\,-\Theta(ck-\omega)\frac{4}{\Ima(k_z)}\Rea\Bigl(\frac{k_{zm}r}{1-r^2}\Bigr)e^{2\phi\Ima(k_z)z},\\
 \mathcal{F}_{0,p}^{\phi,-\phi}(\mathbf{k},\omega)&=-\Theta(\omega-ck)\frac{4}{k_z}\Ima\Bigl(\frac{k_{zm}r}{1-r^2}\Bigr)e^{-2i\phi k_zz}\\
 &\,+\Theta(ck-\omega)\frac{2}{\Ima(k_z)}\Rea\Bigl(k_{zm}\frac{1+r^2}{1-r^2}\Bigr),
\end{split}\end{equation}
which are both diagonal with respect to $\omega$, $\mathbf{k}$, and
$p$ due to the time- and translation-invariance characterizing the
slab. Furthermore, it is proportional to $dz$, and thus goes to zero
in absence of the film. 

Equations~\eqref{corr} and \eqref{fpp0} fully characterize the field
$E_0$ emitted by the film. The counterpropagating components $E^\pm$
of the total field can be expressed as the sum of the individual
contributions of each film, each of which experiences multiple
reflections and transmissions at slab interfaces. The contribution of
a given film of thickness $dz$ reads, 
\begin{equation}\label{sist}
 \begin{cases}
 E^+=u(z_1,z_2)\tau(z_2-z)\Bigl(E_0^++\rho(z-z_1)e^{-2ik_zz}E_0^-\Bigr),\\
 E^-=u(z_1,z_2)\tau(z-z_1)\Bigl(\rho(z_2-z)e^{2ik_zz}E_0^++E_0^-\Bigr), 
 \end{cases}
\end{equation}
where $\rho(\delta)$ and $\tau(\delta)$ are the reflection and
transmission coefficients of a slab of thickness $\delta$ (defined as in~\cite{MessinaPRA14}), and
$u(z_1,z_2)=[1-\rho(z-z_1)\rho(z_2-z)]^{-1}$. In order to deduce the RHT
between the two slabs, we require the correlation functions for
co-propagating components $\langle E_1^+E_1^{+\dag}\rangle$ and
$\langle E_2^-E_2^{-\dag}\rangle$ emitted by the two slabs (see
Fig.~\ref{fig2}). These can be easily obtained from Eqs.~\eqref{corr},
\eqref{fpp0}, and \eqref{sist}. Defining
$\mathcal{F}_{i,p}^{\phi\phi'}(\mathbf{k},\omega)$ for fields $E_i$ produced by each slab ($i=1,2$) as
\begin{equation}\begin{split}
 \langle
 E_{i,p}^\phi(\mathbf{k},\omega)E_{i,p'}^{\phi'\dag}(\mathbf{k}',\omega')&=\frac{\omega}{2\varepsilon_0c^2}(2\pi)^3\delta(\omega-\omega')\\
 &\,\times\delta(\mathbf{k}-\mathbf{k}')\delta_{pp'}\mathcal{F}_{i,p}^{\phi\phi'}(\mathbf{k},\omega),
\end{split}\end{equation}
we obtain:
\begin{widetext}
\begin{equation}\label{fpp}\begin{split}
 \mathcal{F}_1^{++}&=\int_{z_1}^{z_2}\!\!dz\,N[\omega,T(z)]|\tau(z_2-z)u(z_1,z_2)|^2\Bigl[\mathcal{F}_0^{++}+\rho^*(z-z_1)e^{2ik^*_zz}\mathcal{F}_0^{+-}\\
 &\hspace{6cm}+\rho(z-z_1)e^{-2ik_zz}\mathcal{F}_0^{-+}+|\rho(z-z_1)e^{-2ik_zz}|^2\mathcal{F}_0^{--}\Bigr],\\
 \mathcal{F}_2^{--}&=\int_{z_3}^{z_4}\!\!dz\,N[\omega,T(z)]|\tau(z-z_3)u(z_3,z_4)|^2\Bigl[|\rho(z_4-z)e^{2ik_zz}|^2\mathcal{F}_0^{++}+\rho(z_4-z)e^{2ik_zz}\mathcal{F}_0^{+-}\\
 &\hspace{6cm}+\rho^*(z_4-z)e^{-2ik^*_zz}\mathcal{F}_0^{-+}+\mathcal{F}_0^{--}\Bigr],\\
 \end{split}\end{equation}
where for simplicity we have assumed that the two slabs are made of the same material.
 
Following Ref.~\cite{MessinaPRA11}, the flux through a unit area of
the $zz$ component of the Poynting vector in the vacuum region between
the two slabs can be expressed in terms of correlation functions of
the total field $E_T$ between the two slabs as:
\begin{equation}\label{phiS}\begin{split}
\varphi=\frac{1}{(2\pi)^2}\sum_p\int_0^{+\infty}\!\!&d\omega\Biggl[\int_0^{\frac{\omega}{c}}\!\!dk\,k\,k_z\bigl(\mathcal{F}_T^{++}-\mathcal{F}_T^{--}\bigr)+\int_{\frac{\omega}{c}}^{+\infty}\!\!dk\,k\,i\,\mathrm{Im}(k_z)\bigl(\mathcal{F}_T^{+-}-\mathcal{F}_T^{-+}\bigr)\Biggr].
\end{split}\end{equation}
with the total field $E_T$ itself written as the result of multiple
reflections of $E_1^+$ and $E_2^-$ as:
\begin{equation}
 \begin{cases}
 E_T^+=u_{23}\Bigl(E_1^++\rho(z_2-z_1)e^{-2ik_zz_2}E_2^-\Bigr),\\
 E_T^-=u_{23}\Bigl(\rho(z_4-z_3)e^{2ik_zz_3}E_1^++E_2^-\Bigr), 
 \end{cases}
\end{equation}
being $u_{23}=[1-\rho(z_2-z_1)\rho(z_4-z_3)e^{2ik_zd}]^{-1}$. The
total correlation functions are therefore given by:
\begin{equation}\label{FT}
\begin{split}
 \mathcal{F}_T^{++}&=|u_{23}|^2\Bigl(\mathcal{F}_1^{++}+|\rho(z_2-z_1)e^{-2ik_zz_2}|^2\mathcal{F}_2^{--}\Bigr),\\
 \mathcal{F}_T^{+-}&=|u_{23}|^2\Bigl(\rho^*(z_4-z_3)e^{-2ik^*_zz_3}\mathcal{F}_1^{++}+\rho(z_2-z_1)e^{-2ik_zz_2}\mathcal{F}_2^{--}\Bigr),\\
 \mathcal{F}_T^{-+}&=|u_{23}|^2\Bigl(\rho(z_4-z_3)e^{2ik_zz_3}\mathcal{F}_1^{++}+\rho^*(z_2-z_1)e^{2ik_z^*z_2}\mathcal{F}_2^{--}\Bigr),\\
 \mathcal{F}_T^{--}&=|u_{23}|^2\Bigl(|\rho(z_4-z_3)e^{2ik_zz_3}|^2\mathcal{F}_1^{++}+\mathcal{F}_2^{--}\Bigr),
\end{split}
\end{equation}
The above expressions can be simplified in the case of two slabs of
infinite thickness ($z_1\to-\infty$ and $z_4\to+\infty$), in which
case $\rho(\delta)$ becomes the ordinary Fresnel coefficient
$r$. In Eq.~\eqref{phiS} the flux is written as an integral
$\varphi=\int_0^{+\infty}d\omega\,\varphi(\omega)$, with the spectral
components at frequency $\omega=ck_0$ broken down into contributions
from propagative waves $(\omega>ck)$ and evanescent $(\omega<ck)$ waves. Using Eq.~\eqref{FT} and after algebraic manipulations we get the following results for propagative waves
\begin{equation}
\label{phipw}
\varphi_\text{pw}(\omega)=\frac{1}{2\pi^2}\int_0^{k_0}\!\!dk\,k\frac{(1-|r|^2)^2k''_{zm}}{|1-r^2e^{2ik_zd}|^2}\int_0^{+\infty}\!\!\!\!\!dz\,e^{-2k''_{zm}z}\Bigl\{N\Bigl[\omega,T\bigl(-\frac{d}{2}-z\bigr)\Bigr]-N\Bigl[\omega,T\bigl(\frac{d}{2}+z\bigr)\Bigr]\Bigr\},
\end{equation}
and for evanescent waves
\begin{equation}
\label{phiew}
\varphi_\text{ew}(\omega)=\frac{2}{\pi^2}\int_{k_0}^{+\infty}\!\!\!\!dk\,k\frac{(r'')^2e^{-2k_z''d}k''_{zm}}{|1-r^2e^{-2k_z''d}|^2}\int_0^{+\infty}\!\!\!\!\!dz\,e^{-2k''_{zm}z}\Bigl\{N\Bigl[\omega,T\bigl(-\frac{d}{2}-z\bigr)\Bigr]-N\Bigl[\omega,T\bigl(\frac{d}{2}+z\bigr)\Bigr]\Bigr\},
\end{equation}
where $N(\omega,T)=\hbar\omega/[\exp(\hbar\omega/k_BT)-1]$ denotes the
Planck energy of a thermal oscillator, and $c',c''$ denote the real
and imaginary parts of the complex number $c$. As expected, our
expressions simplify in the limit of uniform temperature, reproducing
the typically derived formulas for RHT~\cite{JoulainSurfSciRep05} (note that in addition to the spatial integral over the temperature profiles, our result differs from the typical RHT formula by the extra factor $k''_{zm}$ in the numerator).
 \end{widetext}

\begin{figure*}[t!]
\centering
\includegraphics[height=4.6cm]{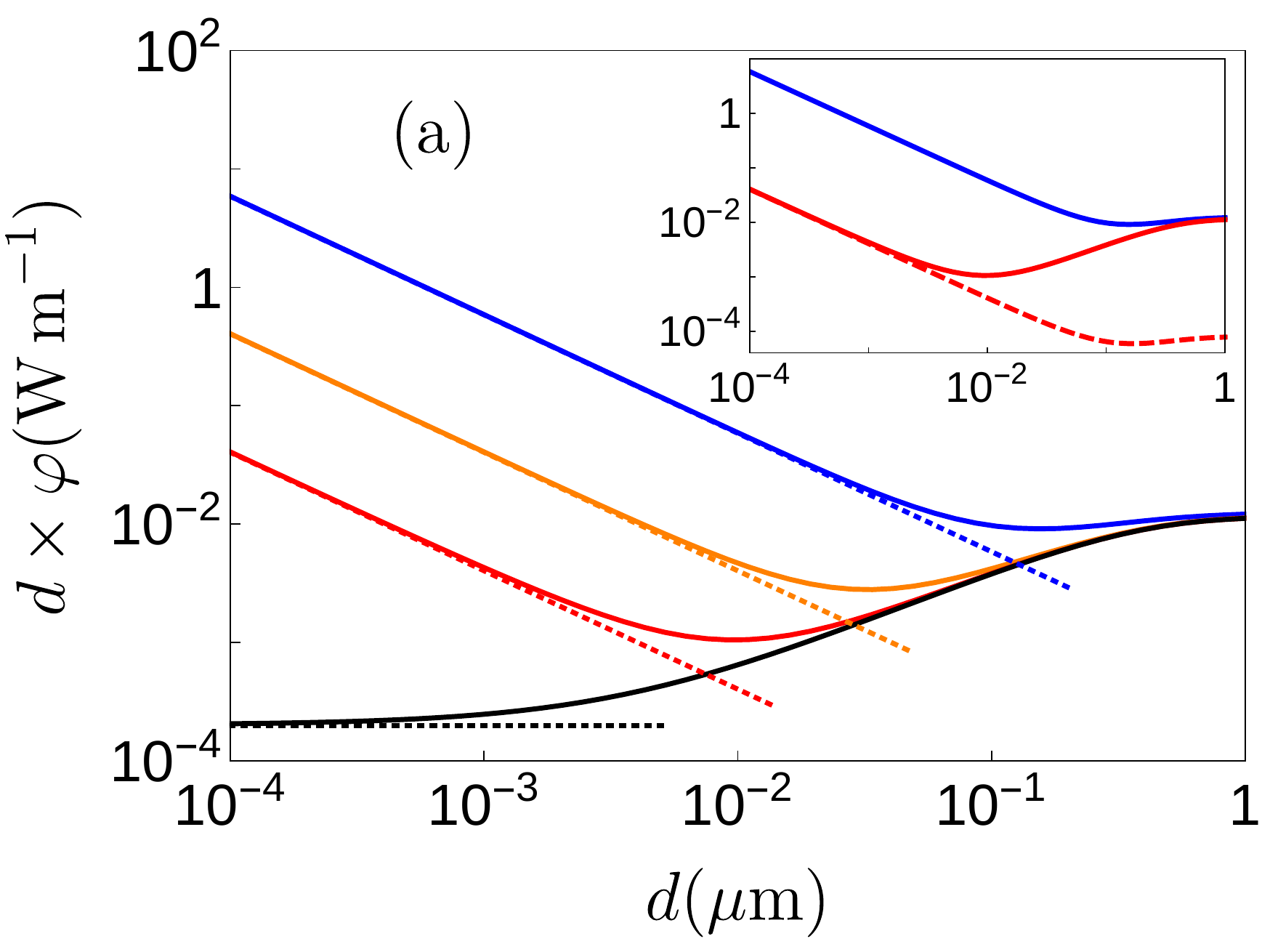}\,\,\,\includegraphics[height=4.6cm]{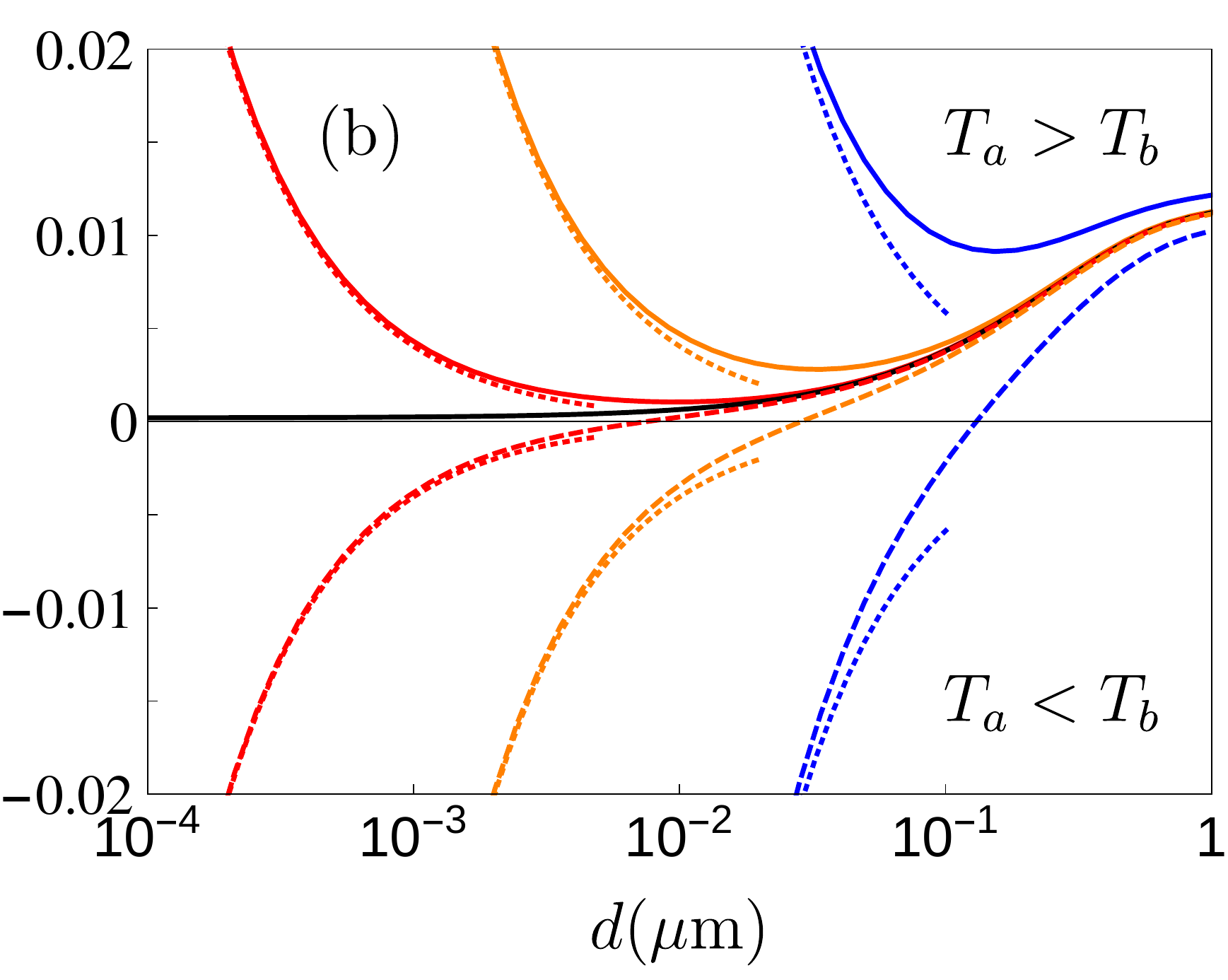}\,\,\,\includegraphics[height=4.6cm]{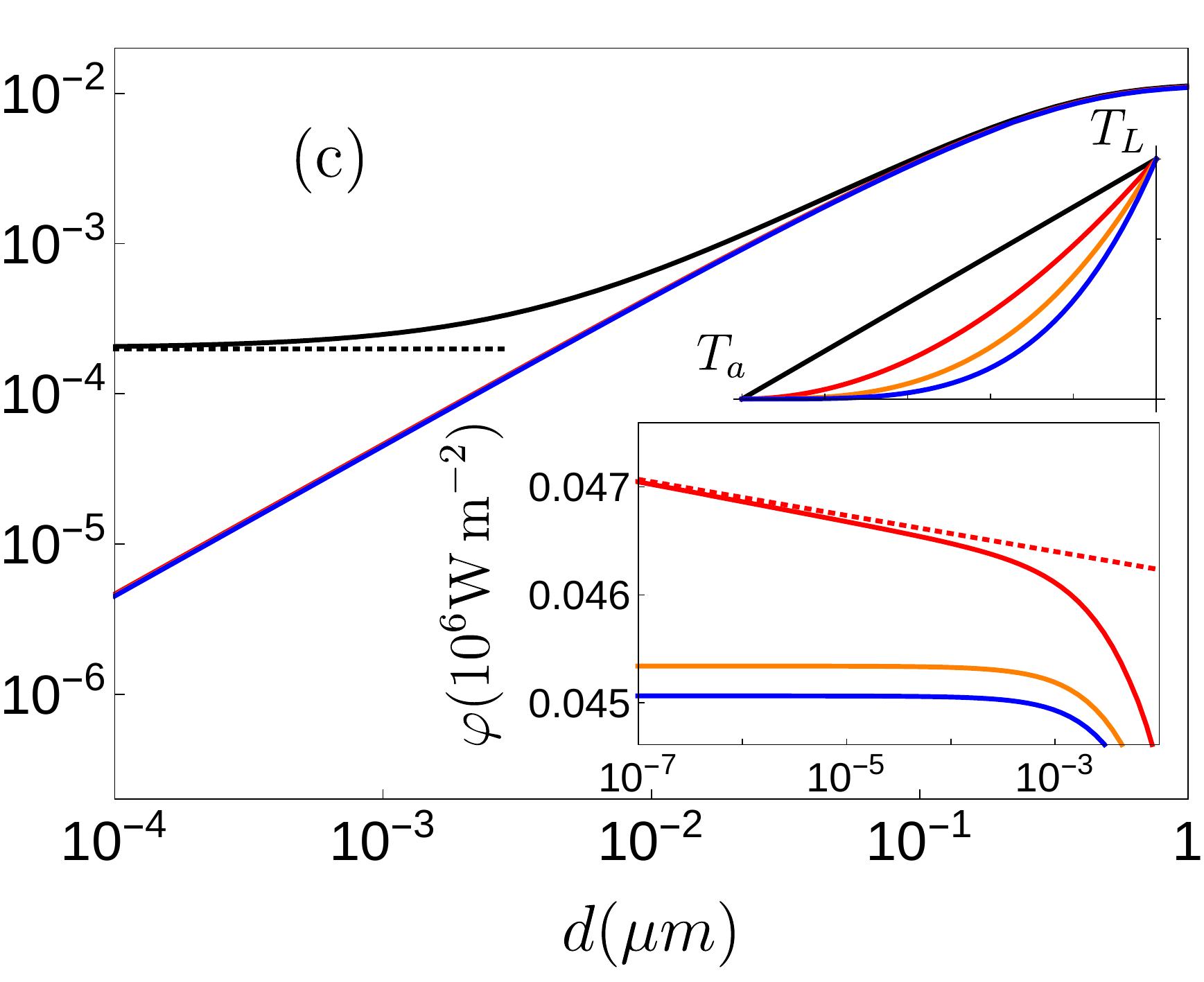}
\caption{(a) Heat flux (multiplied by $d$) between two SiC slabs,
  shown schematically in Fig.~\ref{fig1}(b), separated by a distance
  $d$ and subject to several temperature configurations. In all cases,
  $a=b=1\,\mu$m and $(T_L,T_R)=(600,300)\,$K. Top to bottom plots
  correspond to $(T_a,T_b)=(600,300)\,$K (blue), $(460,440)\,$K
  (orange), $(451,449)\,$K (red) and $(450,450)\,$K (black). Also
  shown are the predictions (dot-dashed lines) of the asymptotic
  formulas given in \eqref{phi0} and \eqref{phi1}. The inset compares
  the flux corresponding to the case
  $(T_L,T_a,T_b,T_R)=(600,451,449,300)\,$K against two
  uniform-temperature configurations $(600,600,300,300)\,$K (solid
  blue line) and $(451,451,449,449)\,$K (dashed red line). (b) Same
  curves in (a) but plotted in a linear scale, with the addition of
  three configurations (dashed lines): $(T_a,T_b)=(449,451)\,$K (red),
  $(440,460)\,$K (orange) and $(300,600)\,$K (blue). (c) Heat flux
  associated with the temperature configuration
  $(T_L,T_a,T_b,T_R)=(600,450,450,300)\,$K but under different
  temperature variations (top inset), described as polynomials of
  orders $n$, corresponding to linear ($n=1$, black), quadratic
  ($n=2$, red), cubic ($n=3$, orange), or quartic ($n=4$, blue)
  polynomials. The black dot-dashed line shows the asymptotic behavior
  predicted by~\eqref{phi1}. The bottom inset illustrates the
  asymptotic behavior of the $n>1$ profiles, with the red dot-dashed
  line corresponding to the prediction of~\eqref{phi2}.}
\label{fig3}
\end{figure*}

\section{Asymptotic behavior}

We are interested in studying the impact of temperature gradients in
the asymptotic limit $d\to0$, in which case RHT is dominated by
evanescent contributions from the transverse-magnetic
polarization. Taylor expanding the population functions around the
slab--vacuum interfaces,
\begin{equation}
 N\Bigl[\omega,T\bigl(\pm\frac{d}{2}\pm
   z\bigr)\Bigr]=\sum_{n=0}^{+\infty}\frac{\alpha^\pm_n(\omega)}{n!}z^n,
\end{equation}
we obtain the RHT $\varphi(\omega)=\sum_n\varphi_n(\omega)$ in
increasing orders of the temperature away from the interface, with
\begin{equation}
\begin{split}
&\varphi_n(\omega)=\frac{\alpha^-_n(\omega)-\alpha^+_n(\omega)}{2^n\pi^2}\int_{k_0}^{+\infty}\!\!\!\!\frac{k\,dk}{(k''_{zm})^n}\frac{(r'')^2e^{-2k''_zd}}{|1-r^2e^{-2k''_zd}|^2}.
\end{split}
\end{equation}
Since the integrand behaves as $k^{1-n}$, it follows that terms of
order $n\geq3$ contribute finite RHT whereas those of order
${n=0,1,2}$ diverge in the limit $d\to0$. Such a divergence is
associated with the increasing contribution of large-$k$ states,
allowing us to approximate the integral. In this limit,
$k''_{zm}\simeq k$, $r$ approaches the $k$-independent quantity
$(\varepsilon(\omega)-1)/(\varepsilon(\omega)+1)$, and it is possible
to take the limit $k_0 \to 0$, allowing us to perform the various $k$
integrals explicitly. Specifically, performing the change of variable
$x=e^{-2kd}$, we obtain:
\begin{equation}\label{phi0}
\varphi_0(\omega)\simeq\frac{1}{8\pi^2d^2}\frac{r''}{r'}\Ima[\text{Li}_2(r^2)][\alpha^-_0(\omega)-\alpha^+_0(\omega)],
\end{equation}
where
$\alpha^-_0(\omega)-\alpha^+_0(\omega)=N(\omega,T_a)-N(\omega,T_b)$
and $\text{Li}_n(z)=\sum_{k=1}^{+\infty}z^k/k^n$ is the
polylogarithmic function. Hence, one finds that to zeroth order in the gradient expansion at the interface, the RHT $\sim 1/d^2$ as $d\to
0$ whenever $T_a \neq T_b$. In contrast, if the temperatures at the
interfaces coincide, this divergence is regularized and the leading
contribution instead comes from the $n=1$ term, given by:
\begin{equation}\label{phi1}
  \varphi_1(\omega)\simeq-\frac{1}{8\pi^2d}\frac{r''}{r'}\Ima[\log(1-r^2)][\alpha^-_1(\omega)-\alpha^+_1(\omega)].
\end{equation}
where, assuming $T_a=T_b$, one finds that
$\alpha^-_1(\omega)-\alpha^+_1(\omega)=-[\partial_zT(-d/2)+\partial_zT(d/2)]\partial_T
N(\omega,T_a)$ depends on the derivatives of the temperature profile
at $z=\pm d/2$. It follows that if $\partial_z T(-d/2)\neq
-\partial_zT(d/2)$ and $T_a=T_b$, the asymptotic behavior of the RHT
$\sim 1/d$. If the former is violated, e.g. when the profile has
zero derivative at the interfaces, then the $n=1$ term is exactly
zero, and the asymptotic behavior is instead determined by the $n=2$
term, which requires a more delicate treatment. In particular,
replacing the integrand by its high-$k$ behavior and performing a
different change of variables $x=\log(ck/\omega)/\log(\omega d/c)$,
one finds:
\begin{multline}
\varphi_2(\omega)\simeq-\frac{[\Ima(r)]^2}{4\pi^2}\log\Bigl(\frac{\omega
    d}{c}\Bigr)[\alpha^-_2(\omega)-\alpha^+_2(\omega)] \\ \times\int_{-\infty}^0dx\,\frac{e^{-2k(x)d}}{|1-r^2e^{-2k(x)d}|^2},
\end{multline}
with $k(x)\equiv k_0\exp[\log(\omega d/c)x]$. We now observe that as
$d\to 0$, the function $\exp[-2k(x)d]$ tends to 1 for any ${-1<x<0}$ and
to 0 for any $x<-1$. Thus, if $T_a=T_b$ and
$\partial_zT(-d/2)=-\partial_zT(d/2)$, it follows that
\begin{equation}\label{phi2}
\varphi_2(\omega)\simeq-\frac{1}{4\pi^2}\log\Bigl(\frac{\omega d}{c}\Bigr)\frac{(r'')^2}{|1-r^2|^2}[\alpha^-_2(\omega)-\alpha^+_2(\omega)].
\end{equation}
where
$\alpha^-_2(\omega)-\alpha^+_2(\omega)=[\partial^2_zT(-d/2)-\partial^2_zT(d/2)]\partial_TN(\omega,T_a)$
involves only second derivatives of $T(z)$ at $z=\pm d/2$. Such a
logarithmic divergence is further regularized if
$\partial^2_zT(-d/2)=\partial^2_zT(d/2)$, in which case the RHT tends
to a constant value in the limit $d\to0$.  A trivial situation under
which all three conditions lead to constant flux as $d\to 0$ is an even temperature profile, i.e. $T(-z) = T(z)$, in which case the flux vanishes at every $d$.

\section{Numerical predictions}

In order to discuss the rich scenarios associated with the presence of temperature gradients, we
consider numerical evaluation of the above formulas for the case of
two infinitely thick parallel silicon carbide (SiC) slabs separated by
vacuum. We consider the specific configuration depicted in
Fig.~\ref{fig1}(b), in which the temperature of the slab on the left
(right) is constant and equal to $T_L$ ($T_R$) everywhere except for a
region of thickness $a$ ($b$), with $T_a$ ($T_b$) denoting the
slab--vacuum interface temperatures of the left (right) slab. Such a
scenario would arise, for instance, if both slabs were to be connected
to thermal reservoirs held at $T_L$ and $T_R$. The dielectric
properties of SiC are described by means of a Drude-Lorenz model
\cite{Palik98}, highlighting the existence of a surface
phonon-polariton resonance in the infrared region of the spectrum,
particularly relevant for near-field RHT
\cite{JoulainSurfSciRep05}. We fix $a=b=1\,\mu$m, focusing first on
the case $(T_L,T_R)=(600,300)\,$K and assuming a linear temperature
gradient in the regions of varying temperature, determined by our
choice of $T_a$ and $T_b$.

Figure~\ref{fig3}(a) shows the RHT (multiplied by $d$) over a wide
range of $d \in [10^{-4},1]\,\mu$m. Note that we include extremely low
values of separations (below a nanometer) in order to better
illustrate the asymptotic regimes discussed above. We consider three
configurations in which $T_a$ differs from $T_b$, illustrating the
expected $1/d^2$ scaling described by~\eqref{phi0}, plotted as dotted
lines, the appearance of which depends on the precise values of $T_a$
and $T_b$, with the transition occuring anywhere between a few to
hundreds of nm. Also shown is the RHT in the special case
${T_a=T_b=450\,}$K, illustrating the $1/d$ behavior predicted
by~\eqref{phi1} (dotted line), the onset of which occurs below the nm
scale. Noticeably, while all four curves approach one another at the
micron scale, the different values of interface temperatures produce
both quantitatively and qualitatively different behaviors in the
experimentally accessible range ${d \in [1,100]\,}$nm. It is instructive
to compare one of the above configurations,
$(T_L,T_a,T_b,T_R)=(600,451,449,300)\,$K, to the more standard
scenario of uniform-temperature slabs: $(600,600,300,300)\,$K and
$(451,451,449,449)\,$K. The results, shown in the inset of
Fig.~\ref{fig3}(a), demonstrate that at small distances, RHT becomes a
surface effect, in which case only the interface temperatures are
relevant; in contrast, at large $d$ RHT is dominated (and well
described) by the bulk temperatures $T_L$ and $T_R$ of the infinite
regions.

Figure~\ref{fig3}(b) shows the four curves of Fig.~\ref{fig3}(a) in a
linear scale and introduces three additional configurations,
corresponding to situations in which $T_a \leftrightarrow T_b$ are
exchanged (dashed lines). Such a flip leads to a situation in which
the bulk ($T_L > T_R$) and surface ($T_a < T_b$) temperatures compete,
contributing RHT in opposite directions. As before, the behavior at
asymptotically small $d$ is determined by~\eqref{phi0} and
\eqref{phi1} (dotted lines), except that in the case of flipped $T_a <
T_b$ (dashed lines), the RHT goes from positive to negative (reversing
sign) as $d$ decreases, with the transition distance occuring anywhere
from a few to hundreds of nm, depending on $T_a,T_b$. Such a surface-temperature inversion could potentially be engineered (and tuned) via the introduction of an external pump or thermostat.

\begin{figure}
\begin{center}
\includegraphics[height=6.1cm]{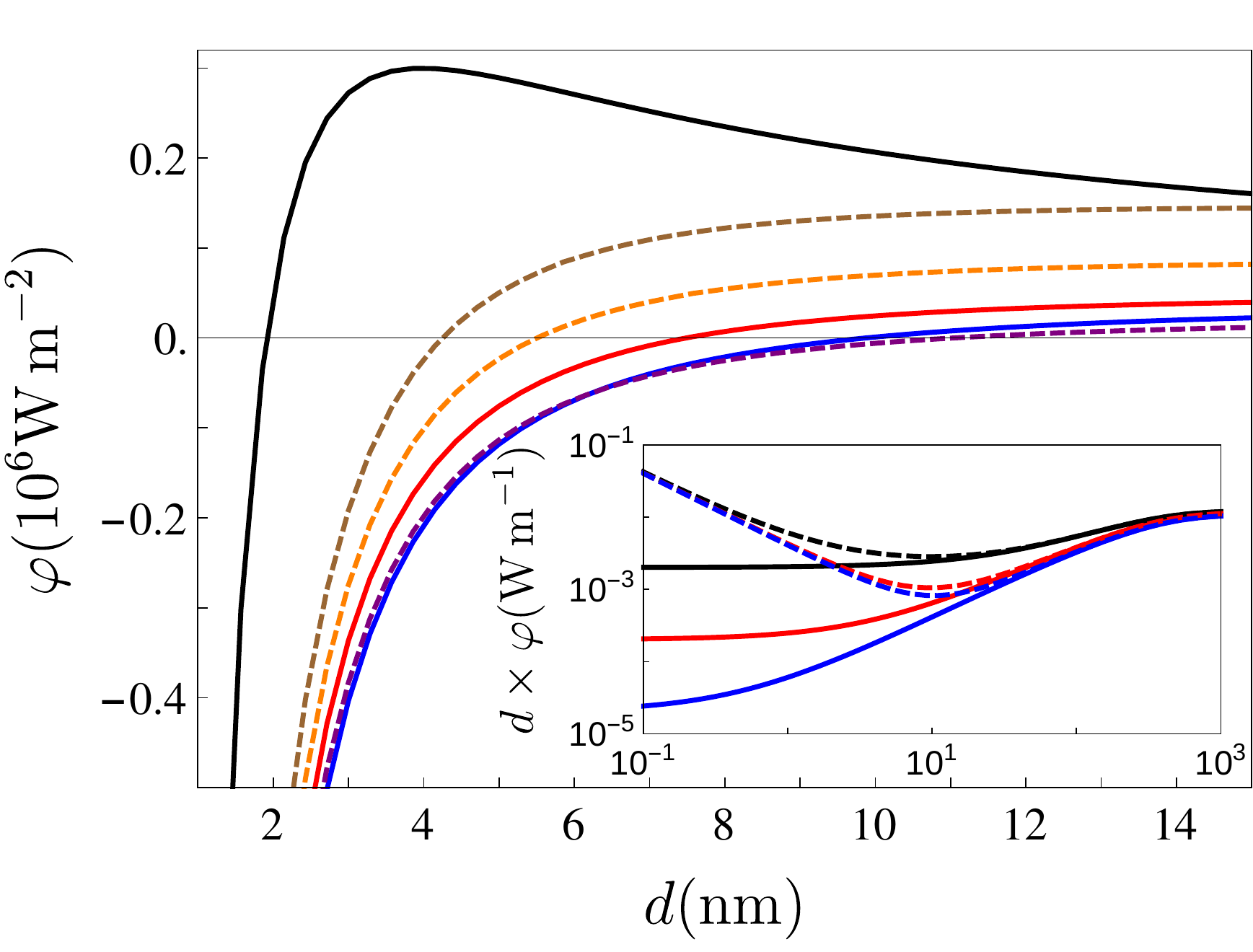}
\end{center}
\caption{Heat flux as a function of distance for
  $(T_a,T_b,T_L)=(451,449,300)\,$K. The solid lines correspond to
  $T_L=600\,$K and have $a=b=100\,$nm (black), $1\,\mu$m (red) and
  $10\,\mu$m (blue). The dashed lines correspond to $a=b=1\,\mu$m and
  have $T_L=800\,$K (brown), $700\,$K (orange) and $500\,$K (purple).}
\label{fig4}
\end{figure}

The results presented thus far highlight the existence of both
$1/d^2$ and $1/d$ asymptotic power-laws. Figure~\ref{fig3}(c) on the
other hand also illustrates the appearance of logarithmic behavior by
considering a situation consisting of fixed
$(T_L,T_a,T_b,T_R)=(600,450,450,300)\,$K but where the intervening
temperature profile is chosen to have different polynomial
dependencies (shown on the top inset), including linear ($n=1$,
black), quadratic ($n=2$, red), cubic ($n=3$, orange), and quartic
($n=4$, blue) power laws. While the sub-$1/d$ behavior associated
with the $n>1$ profiles is apparent from the main plot, the three
curves are better distinguished in the inset of the figure, which
shows the slow, logarithmic scaling associated with the $n=2$ profile,
plotted in conjunction with the predictions of~\eqref{phi2} (dotted
line), along with the fact that RHT approaches a constant for $n>2$.

Figure~\ref{fig4} focuses on the role of the thicknesses ${a=b}$ and
external temperature $T_L$ on the sign-flip effect explored in
Fig.~\ref{fig3}(b), considering the reference scenario
$(T_L,T_a,T_b,T_R)=(600,451,449,300)\,$K. We first fix ${T_L=600\,}$K
and vary the thickness, from $a=100$\,nm to $a=10\,\mu$m (black, red, and
blue lines), demonstrating a decreasing zero-flux distance, from 10nm
to 2nm, with decreasing thickness. Fixing $a=1\,\mu$m and modifying
instead the external temperature, from $T_L=500\,$K to ${T_L=800\,}$K,
produces similar variations on the zero-flux distance, from 4nm to
10nm. The inset of Fig.~\ref{fig4} yields even more insights on ways
of manipulating the asymptotic behavior, showing the RHT (multiplied
by $d$) for the same three values of
${a=100\,\text{nm},1\,\mu\text{m},10\,\mu\text{m}}$ explored in the main
figures, but under different surface temperatures (or gradients). The
dashed lines correspond to the case $(T_a,T_b)=(451,449)\,$K,
illustrating the expected $1/d^2$ scaling behavior. It follows
from~\eqref{phi0} that in this case the asymptotic RHT depends only on
the two temperatures $T_a$ and $T_b$ and not on their derivatives,
which explains why the three dashed lines approach one another as $d
\to 0$. The solid lines correspond to the case $T_a=T_b=450\,$K and
illustrate the expected $1/d$ behavior, revealing an asymptotic
prefactor that decreases with decreasing temperature gradients, as
predicted by~\eqref{phi1}.

\section{Conclusions}

The approach we presented, valid for arbitrary materials and distances and based on a scattering-matrix formalism, leads to analytical expressions of the short-distance behavior of the flux. We have shown that the latter is entirely determined by the gradient expansion of the temperature profile near the slab–vacuum interfaces. In particular, we find that apart from the well-known $1/d^2$ power-law scaling, under certain conditions, the flux can diverge asymptotically either as $1/d$ or logarithmically, or it can also saturate to a constant value. We have shown that the introduction of a temperature profile can result in significant flux tunability, leading for instance to changes in the sign of the flux with respect to slab separations. The temperature profile within a given body can be for example experimentally engineered by means of the introduction of several thermostats put in contact at different points of the body. Moreover, a temperature gradient can naturally appear as the result of the coupling between radiative exchange and conduction within each body, as studied in detail in Refs.~\cite{MessinaPRB,JinarXiv}, in both planar and structured geometries. It has been shown that, depending on the chosen material, an observable temperature profile can indeed appear for distances as high as tens or hundreds of nanometers. Our approach would be needed to accurately describe radiative heat transfer under these conditions. In fact, recent experiments are beginning to explore such short distance regime (down to sub-nanometer separations~\cite{KittelPRL05,KloppstecharXiv,ShenNanoLetters09,KimNature15}), some of which have already observed a saturating flux that has yet to be properly explained. Moreover, the possibility of tuning the temperature profile of a system and thereby the behavior, e.g. sign, of the heat transfer with respect to external thermal sources is yet unexplored and could be important for thermal devices~\cite{devices}. Finally, it must be stressed that at distances as low as a few nanometers or in the sub-nanometer range theoretical descriptions based on macroscopic fluctuational electrodynamics are no longer valid: in this regime, atomic-scale and other non-local screening effects as well as the tunneling of phonons can play significant role.

\section*{Acknowledgements} This work was supported by the National Science Foundation under Grant no. DMR-1454836 and by the Princeton Center for Complex Materials, a MRSEC supported by NSF Grant DMR
1420541. We thank M. Kr\"{u}ger for pointing out the similarities that arise in the description of near-field heat transfer between rough, curved surfaces and objects under temperature gradients (our work).


\begin{thebibliography}{99}
\bibitem{JoulainSurfSciRep05}K. Joulain \emph{et al.}, Surf. Sci. Rep. \textbf{57}, 59 (2005).
\bibitem{ChapuisPRB08}P.-O. Chapuis \emph{et al.}, Phys. Rev. B \textbf{77}, 035431 (2008).
\bibitem{MuletMTE02}J.-P. Mulet, K. Joulain, R. Carminati, and J.-J.Greffet, Microscale Thermophysical Engineering \textbf{6}, 209 (2002).
\bibitem{NarayanaswamyPRB08}A. Narayanaswamy, S. Shen, and G. Chen, Phys. Rev. B \textbf{78}, 115303 (2008).
\bibitem{HuApplPhysLett08}L. Hu, A. Narayanaswamy, X. Chen, and G. Chen, Appl. Phys. Lett. \textbf{92}, 133106 (2008).
\bibitem{ShenNanoLetters09}S. Shen, A. Narayanaswamy, and G. Chen, Nano Letters \textbf{9}, 2909 (2009).
\bibitem{RousseauNaturePhoton09}E. Rousseau, A. Siria, G. Joudran, S. Volz, F. Comin, J. Chevrier, and J.-J. Greffet, Nature Photon. \textbf{3}, 514 (2009).
\bibitem{OttensPRL11}R. S. Ottens, V. Quetschke, S. Wise, A. A. Alemi, R. Lundock, G. Mueller, D. H. Reitze, D. B. Tanner, and B. F. Whiting, Phys. Rev. Lett. \textbf{107}, 014301 (2011).
\bibitem{KralikRevSciInstrum11}T. Kralik, P. Hanzelka, V. Musilova, A. Srnka, and M. Zobac, Rev. Sci. Instrum. \textbf{82}, 055106 (2011).
\bibitem{KralikPRL12}T. Kralik, P. Hanzelka, M. Zobac, V. Musilova, T. Fort, and M. Horak, Phys. Rev. Lett. \textbf{109}, 224302 (2012).
\bibitem{vanZwolPRL12a}P. J. van Zwol, L. Ranno, and J. Chevrier, Phys. Rev. Lett. \textbf{108}, 234301 (2012).
\bibitem{vanZwolPRL12b}P. J. van Zwol, S. Thiele, C. Berger, W. A. de Heer, and J. Chevrier, Phys. Rev. Lett. \textbf{109}, 264301 (2012).
\bibitem{SongNatureNano15}B. Song \emph{et al.}, Nature Nanotechnology \textbf{10}, 253 (2015).
\bibitem{KimNature15}K. Kim \emph{et al.}, Nature \textbf{528}, 387 (2015).
\bibitem{StGelaisNatureNano16}R. St-Gelais, L. Zhu, S. Fan, and M. Lipson, Nature Nanotechnology \textbf{11}, 515 (2016).
\bibitem{KittelPRL05}A. Kittel, W. M\"{u}ller-Hirsch, J. Parisi, S.-A. Biehs, D. Reddig, and M. Holthaus, Phys. Rev. Lett. \textbf{95}, 224301 (2005).
\bibitem{KloppstecharXiv}K. Kloppstech \emph{et al.}, preprint arXiv:1510.06311 (2015).
\bibitem{Rodriguez13} A. W. Rodriguez, M. T. H. Reid, J. Varela, J. D. Joannopoulos, F. Capasso, and S. G. Johnson, Phys. Rev. Lett. \textbf{110}, 014301 (2015).
\bibitem{HenkelApplPhysB06}C. Henkel and K. Joulain, Appl. Phys. B \textbf{84}, 61 (2006).
\bibitem{JoulainJQSRT}K. Joulain, J. Quant. Spectrosc. Radiat. Transfer \textbf{109}, 294 (2008).
\bibitem{ChiloyanNatComm15}V. Chiloyan, J. Garg, K. Esfarjani, and G. Chen, Nature Comm. \textbf{6}, 6755 (2015).
\bibitem{Kruger}M. Kr\"{u}ger, V. A. Golyk, G. Bimonte, and M. Kardar, Europhys. Lett. \textbf{104}, 41001 (2013).
\bibitem{MessinaPRB}R. Messina, W. Jin, and A. W. Rodriguez, Phys. Rev. B \textbf{94}, 121410(R) (2016).
\bibitem{JinarXiv}W. Jin, R. Messina, and A. W. Rodriguez, preprint arXiv:1605.05708 (2016).
\bibitem{FVC}A. G. Polimeridis, M. T. H. Reid, W. Jin, S. G. Johnson, J. K. White, and A. W. Rodriguez, Phys. Rev. B \textbf{92},134202 (2015). 
\bibitem{Eda1}S. Edalatapour and M. Francoeur, J. Quant. Spectrosc. Radiat. Transfer \textbf{133}, 364 (2014).
\bibitem{Eda2}S. Edalatapour and M. Francoeur, Phys. Rev. B \textbf{94}, 045406 (2016).
\bibitem{devices}P. Ben-Abdallah and S.-A. Biehs, AIP Advances \textbf{5}, 053502, (2015).
\bibitem{rect}L. Zhu, C. R. Otey, and S. Fan, Appl. Phys. Lett. \textbf{100},  044104 (2012).
\bibitem{MessinaPRA11}R. Messina and M. Antezza, Phys. Rev. A \textbf{84}, 042102 (2011).
\bibitem{MessinaPRA14}R. Messina and M. Antezza, Phys. Rev. A \textbf{89}, 052104 (2014).
\bibitem{Palik98}\emph{Handbook of Optical Constants of Solids}, edited by E. Palik (Academic Press, New York, 1998).
\end{thebibliography}
\end{document}